\documentclass[
 aip,
 amsmath,amssymb,
 reprint,%
]{revtex4-1}

\usepackage{graphicx}
\usepackage{dcolumn}
\usepackage{bm}

\usepackage[utf8]{inputenc}
\usepackage[T1]{fontenc}
\usepackage{mathptmx}
\usepackage{etoolbox}

\makeatletter
\def\@email#1#2{%
 \endgroup
 \patchcmd{\titleblock@produce}
  {\frontmatter@RRAPformat}
  {\frontmatter@RRAPformat{\produce@RRAP{*#1\href{mailto:#2}{#2}}}\frontmatter@RRAPformat}
  {}{}
}%
\makeatother
\begin{document}

\preprint{AIP/123-QED}

\title{The Effect of Hydration and Dynamics on the Mass Density of Single Proteins}

\author{C.C.W. McAllister}
\affiliation{Department of Physics, Durham University, Durham, UK}

\author{L.S.P. Rudden}
\affiliation{Institute of Bioengineering, École Polytechnique Fédérale de Lausanne, Lausanne, Switzerland}

\author{E.H.C. Bromley}
\email{e.h.c.bromley@durham.ac.uk}
\affiliation{Department of Physics, Durham University, Durham, UK}

\author{M.T. Degiacomi}
\email{matteo.degiacomi@ed.ac.uk}
\affiliation{Department of Physics, Durham University, Durham, UK}
\affiliation{EaStCHEM School of Chemistry, University of Edinburgh, Edinburgh, UK}
\affiliation{School of Informatics, University of Edinburgh, Edinburgh, UK}

\date{\today}

\begin{abstract}
\section{Abstract}
\noindent

The density of a protein molecule is a key property within a variety of experimental techniques. We present a computational method for determining protein mass density that explicitly incorporates hydration effects. Our approach uses molecular dynamics simulations to quantify the volume of solvent excluded by a protein. Applied to a dataset of 260 soluble proteins, this yields an average density of 1.296 $\pm$ 0.001 g cm\textsuperscript{-3}, notably lower than the widely cited value of 1.35 g cm\textsuperscript{-3}. Contrary to previous suggestions, we find no correlation between protein density and molecular weight. We instead find correlations with residue composition, particularly with hydrophobic amino acid content. Using these correlations, we train a regressor capable of accurately predicting protein density from sequence-derived features alone. Examining the effect of incorporating water molecules on the measured density, we find that water molecules buried in internal cavities have a negligible effect, whereas those at the surface have a profound impact. Furthermore, by calculating the density of a titin domain and of the Bovine Pancreatic Trypsin over molecular dynamics trajectories, we show that individual proteins can occupy states with close but distinguishable densities. Finally, we analyse the density of water in the vicinity of proteins, showing that the first two hydration shells exhibit higher density than bulk water. When included in cumulative density calculations, these hydration layers contribute to a net increase in local solvent density. Overall, we find that proteins are less dense than previously reported, which is offset by their ability to induce a higher density of water in their vicinity.
\end{abstract}


\maketitle 

\section{Introduction}
\noindent Protein mass density --- the density of an individual protein molecule --- is an important fundamental biophysical quantity relevant to a wide range of experimental techniques, including X-ray crystallography and ultracentrifugation studies of protein oligomers~\cite{Chetri2022}. Additionally, the precise determination of protein and solvent densities is relevant to methods for which the local dielectric permittivity is experimentally implicated, including Extraordinary Acoustic Raman Spectroscopy (EARS) ~\cite{wheaton2015probing}. 
The protein mass density has been approximated since at least the late 1960s ~\cite{Matthews1968} to be effectively a constant independent of the protein’s size, shape, or other physical characteristics, a consequence of the closely packed interiors of proteins ~\cite{Erickson2009}. Specifically, a value of 1.35 g cm\textsuperscript{-3} has been commonly used ~\cite{Andersson1998} based on early compressibility ~\cite{Gekko1979} and sedimentation ~\cite{Squire1979} studies, though various values deriving from experimental and theoretical approaches have been proposed. \par

Key to the calculation of protein density is the calculation of the protein volume. There are many different definitions of protein volume, including: the geometric volume, which is the solvent-excluded volume enclosed within the solvent-excluded surface ~\cite{Greer1978}, also known as the molecular surface volume ~\cite{Chen2015}; the van der Waals volume, the volume of overlapping spheres representing the van der Waals radii of each constituent atom of the protein ~\cite{Lee1971}; and the solvent-accessible volume, the volume enclosed by the solvent-accessible surface area (SASA) ~\cite{Richmond1984}. This work is concerned with the molecular surface volume. This is the relevant measure of volume occupied by a protein for experiments which depend on changes in the local permittivity of proteins in solvent (e.g., EARS).\par

While protein mass density is often taken as a constant, in 2004 Fischer \textit{et al.} 
 ~\cite{Fischer2004} examined previously published experimental ~\cite{Squire1979, Gekko1979} and theoretical ~\cite{Quillin2000, Tsai1999, Andersson1998} estimates of protein densities and concluded that for relatively small proteins (below 20 kDa) protein density is molecular weight dependent, in an inverse exponential relationship. The authors further proposed that, for larger proteins, a constant value of 1.410(6) g cm\textsuperscript{-3} should be considered. 
Theoretical calculations of protein density used in the meta-analysis of Fischer \textit{et al.} were performed on crystal structures almost completely devoid of water molecules using Voronoi Tessellation methods (Andersson and Hovmöller, 1998 ~\cite{Andersson1998}; Tsai \textit{et al.}, 1999 ~\cite{Tsai1999}; Quillin and Matthews, 2000 ~\cite{Quillin2000}). These techniques provide analogous approximations of a dry molecular surface volume. Crucially though, water molecules form hydration shells around proteins, whereby coordinated waters essentially behave as an integral part of the protein ~\cite{Vinh2011}. The specific distribution of conformational states of protein side chains depends on their complex interactions with these water molecules, thus the evaluation of protein-water dynamics is required to accurately calculate protein volume. Water-protein interactions play key structural roles in proteins, driving their organisation and flexibility, and ultimately impacting upon their function 
 ~\cite{Martini2013, Biedermannova2016}. For this reason, computational methods that can accurately incorporate hydration into the calculation of protein density would be advantageous.
Multiple methods to calculate the molecular surface volume exist including analytic methods like MSMS (Michel Sanner’s Molecular Surface), which can compute the molecular surface volume via a procedure that relies on the reduced surface ~\cite{sanner1996}, and explicit surface representation methods like LSMS (Level Set method for Molecular Surface generation), which uses a level-set front-propagation method ~\cite{Can2006}. Inferring protein volumes from their amino acid sequence, with the volume calculated as the sum of the volumes of the constituent amino acids, has also been shown to produce remarkably accurate results ~\cite{Hutt2012}. A recent study of the partial specific volumes of proteins, the inverse of the protein density in solution, using this method has calculated a theoretical mean value for all human proteins of 0.735 mlg\textsuperscript{-1} with a standard deviation of 0.010 mlg\textsuperscript{-1}, equivalent to 1.36 $\pm$ 0.03 g cm\textsuperscript{-3}, and an approximately Gaussian distribution ~\cite{Zhao2011}. However, none of these methods account for the effects of the protein’s hydration shell. Furthermore, the partial specific volume is a macroscopic experimentally observable quantity from which the mass density of individual protein molecules cannot necessarily be simply derived. \par

Previous computational studies of protein mass density that have attempted to account for the surface effects of water have mostly utilised solvent-corrected Voronoi tessellation methods ~\cite{Quillin2000, Andersson1998, Tsai1999}. While useful, the Voronoi method is known to produce inaccurate results for surface atoms which are sparsely surrounded by other atoms, leading to different density results depending on how or if the contributions of these surface atoms are accounted for ~\cite{Quillin2000}. The Voronoi method as applied to proteins is usually also adjusted to account for different atomic radii which can introduce vertex errors, though these can be eliminated by enveloping each Voronoi cell with a hyperbolic surface ~\cite{Goede1997}. While some previous work has used Voronoi methods including water molecules from a simulation, dividing the entire simulation box including solvent into Voronoi polyhedra ~\cite{Gerstein1995}, most have focused on calculating volumes from crystal structures. \par

In this work we consider protein density to be associated with the volume of solvent they displace, whereby the position of water molecules coordinated with the protein is explicitly determined and accounted for using all-atom molecular dynamics (MD) simulations. To this end, we adopt a voxel-based method analysing the position of protein and water atoms in molecular dynamics simulation snapshots. This approach offers key advantages. Firstly, utilising MD simulations allows for volumes and densities to be calculated for many protein conformers, yielding better statistics than when single atomic structures are used. Second, it avoids assumptions that may affect density estimation, e.g., water has homogeneous density everywhere and is ideally packed around every protein atom. Third, it enables characterising how changes in environmental conditions (e.g., temperature, pressure, pH, solvent composition) might affect protein density. Hereafter we present our method, before using it to determine protein density in solution for a large set of proteins and assessing whether this property depends on any physical characteristics. Furthermore, to test the assumption of close internal packing leading to a constant protein mass density value ~\cite{Erickson2009}, we present a method to identify the presence of buried water molecules within the protein interior. \par

We use our method to calculate the density of a dataset of 260 soluble proteins. We then train a random forest regressor (RFR) ~\cite{randomforests} with our calculated densities and a range of structural features, demonstrating that soluble protein density can be predicted from amino acid sequence alone. Finally, we examine how mass density might vary within individual proteins  in both equilibrium and non-equilibrium conditions. To this end, we examine two case studies: Bovine Pancreatic Trypsin Inhibitor (BPTI) and immunoglobin-like (Ig-like) domain of titin. BPTI, the first protein simulated with molecular dynamics ~\cite{karplus1977}, has a well-characterised conformational space thanks to a 1-millisecond unbiased simulation performed by DE Shaw et al. ~\cite{Shaw2010}. Here, we utilise Markov State Modelling (MSM) to divide this simulation into discrete states and show that these feature distinct densities. Titin contributes to the passive elasticity of muscle by acting as a molecular spring ~\cite{Granzier1997}. It is the largest known protein consisting of up to 300 mostly Ig-like domains ~\cite{Optiz2003} which unfold sequentially under the influence of an external stretching force, with the domains refolding upon relaxation ~\cite{Rief1997}. Utilising a steered molecular dynamics (SMD) simulation of a single titin Ig-like domain, we evaluate the evolution of protein mass density over the large-scale conformational change induced by mechanical stress. We find that within the SMD simulation there exists two sub-populations --- divided by secondary structure content --- with different densities. \par

Finally, we extend our density calculation method to the hydration shell surrounding a protein, and use it to investigate the structure of water around each protein in our dataset. In agreement with previous experimental ~\cite{svergun1998protein} and simulation ~\cite{merzel2002first} data, we find that the mean first hydration shell density (1.1 $\pm$ 0.3 g cm\textsuperscript{-3}) is 12\% greater than bulk water. Additionally, we find that the second hydration shell is on average significantly more dense (1.5 $\pm$ 0.2 g cm\textsuperscript{-3}, with a density increase of 54.5\% compared to bulk water. The unexpected extent of this increase in the presence of a protein molecule may explain the discrepancy between experimental and computational estimates of protein density.

\section{Methods}

\subsection{Protein dataset}
\noindent
To study the density of soluble proteins, we took a diverse set of 260 protein monomers featured in the protein-protein docking benchmark 5 ~\cite{Vreven2015}. To accurately measure their density, linked to the volume of water they displace, we aim to explicitly determine how water molecules arrange themselves around each protein atom. To this end, we solvated each protein in a TIP3P water box neutralized with Na\textsuperscript{+} and Cl\textsuperscript{$-$} counterions, and relaxed the resulting system using molecular dynamics (MD) simulations using the GROMACS engine and the Amber ff14SB ~\cite{Maier2015} force field. Proteins were first energy minimized using a steepest descent algorithm until a maximum force of less than 1000 kJ mol$^{-1}$ nm$^{-1}$ was achieved. Then, they were all equilibrated in the NVT ensemble at a temperature of 300 K, using a 2 fs timestep with bonds restrained using LINCS. A particle-mesh Ewald ~\cite{Darden1993} summation was used to treat long-range interactions and the velocity-rescaled modified Berendsen temperature coupling method ~\cite{Bussi2007} applied separately to protein and non-protein atoms. Finally, 1 ns production runs were carried out in the NPT ensemble, with 300 K and 1 bar determined by modified Berendsen temperature coupling and Parrinello-Rahman pressure coupling ~\cite{Parrinello1981}. Protein-water conformations were extracted every 50 ps from each production run, leading to the extraction of 5460 snapshots (21 for each protein). Since water density is itself temperature-dependent, to assess the effect of temperature on measured density, we repeated the simulation protocol described above, for all proteins, at a physiological temperature of 310.15 K. Finally, to ensure consistency across force fields and water models, we also repeated our simulation protocol at both 300 and 310.15 K using the Amber ff99-ILDN  ~\cite{amber99ildn} force field and the SPC/E water model (see Supplemental Materials). The specific force field and water model combinations used in this work were chosen for the reported agreement between protein hydration shell contrasts predicted using them, and experimental small angle X-ray and neutron scattering data ~\cite{Linse2023}. Unless otherwise specified, in the main text we report results obtained at 300 K using the Amber ff14SB force field and the TIP3P water model.

\subsection{Protein volume calculation}
\begin{figure}[h!]
  \includegraphics[scale=0.4]{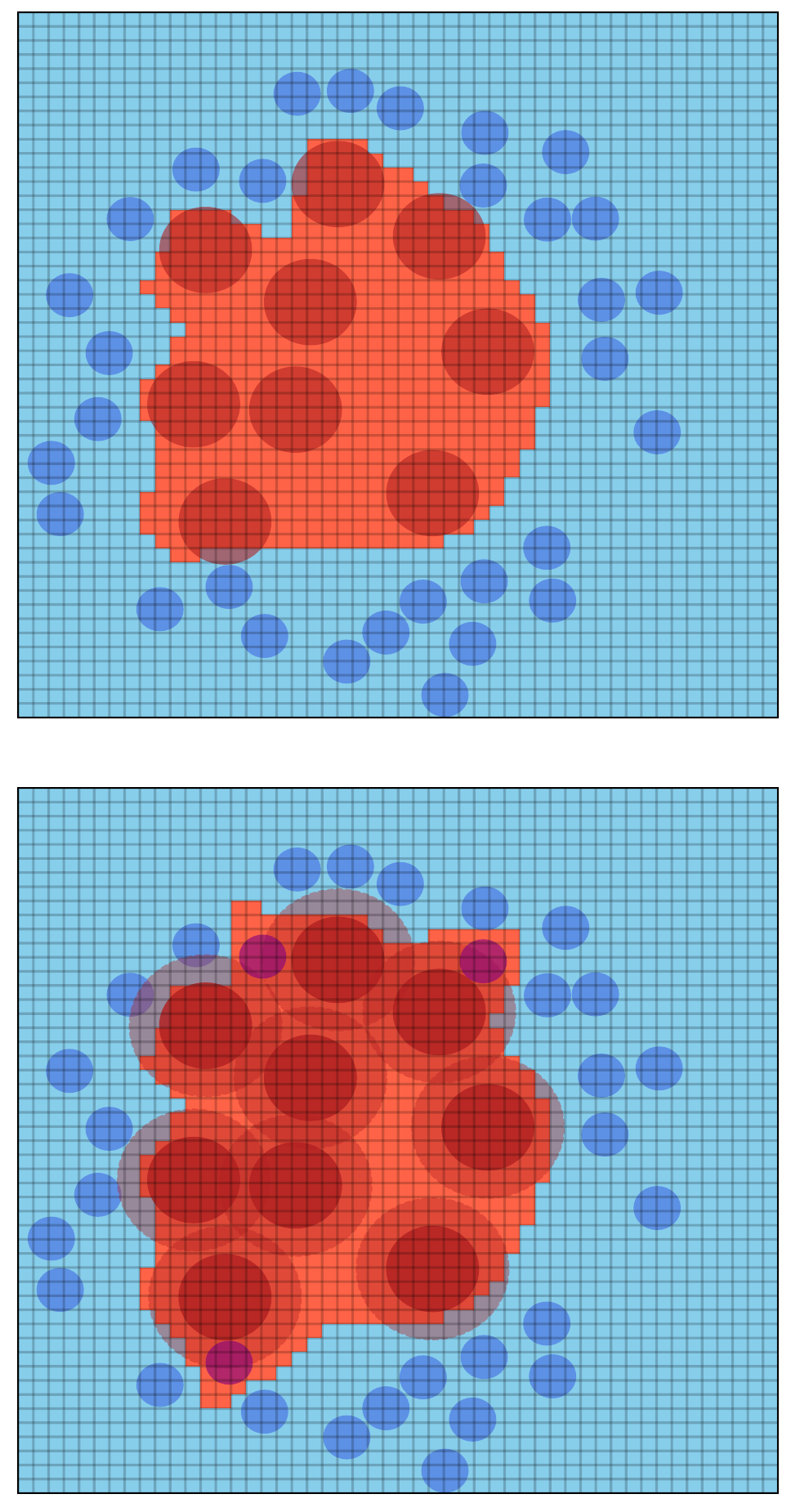}
  \caption{Schematic representation of our protein volume calculation method. Red and blue circles represent the van der Waals radius of protein and water atoms, respectively. The space surrounding the protein is divided in a fine grid, coloured according to which atom is the closest. (top) the red region represents protein occupied volume. (bottom) the red region represents the volume occupied by protein and solvation shell. This is calculated by considering as part of the protein any water atom within a given cutoff distance from the centres of any protein atoms (transparent circles surrounding red circles). This region may be incrementally expanded to allow for the characterisation of changes in water density at different distances from the protein.}
  \label{fig:method}
\end{figure}

\noindent
To calculate the volume occupied by a protein, we place an equilibrated MD simulation snapshot of the protein with its surrounding water into a 3-dimensional grid (Figure \ref{fig:method}) defined by two parameters: the size of its cubic voxels (hereafter “step”), and the amount of extra space added to the grid at the extremities of the protein in each Cartesian axis (“leeway”). Our method is implemented in Python, using NumPy ~\cite{numpy}, SciPy ~\cite{scipy}, and MDAnalysis ~\cite{Gowers2016,Denning2011} packages (see Supplementary Information).
\par

For each layer of the grid corresponding to a specific $Z$-value in Cartesian space we calculate the distances from the centre of each voxel to all protein and solvent atom positions. The distance from the centre of each voxel in the layer to the nearest protein and solvent atoms is then calculated, and every voxel for which the nearest protein atom is closer than the nearest solvent atom is considered part of the protein, with a voxel of volume (simply equivalent to the step size cubed) added to the total calculated volume. By performing the calculation layer-by-layer the memory requirements of the procedure is drastically reduced (see Figures \ref{fig:method} and S1). To make computation more efficient a ‘shell’ parameter was added to define a shell around the protein (or a distance from each protein atom) outside of which water molecules would be excluded from the calculation. We found that a shell distance of 6 Å increases computational efficiency without sacrificing accuracy.
\par
To calculate the density of a protein combined with adjacent water molecules within a cutoff distance, we adapted our algorithm to enable considering selected water molecules as part of the protein itself. To assess how water is distributed around a protein, we calculate the radial distribution function (RDF) between all protein atoms, and all solvent ones. The locations of the first and second minima in the RDF are identified to give the thickness of the first and second hydration shells ~\cite{Liu2021}. \par
Our density calculation algorithm can accommodate the usage of atoms' van der Waals radii, by subtracting the atomic radii from the calculated distances prior to evaluating whether a grid point is closer to protein or water atoms. We assessed the effect of this improvement by either using radii from \textit{A. Bondi, 1964} ~\cite{Bondi1964}, or from the Amber ff14SB force field. Overall, we found that accounting for van der Waals radii has a minimal effect on the calculated volumes. Therefore, to reduce computational time, we consider all atoms as having the same radius (see Figures S6 and S7). \par
The main parameters for the protein volume calculation are the step size of the grid, the leeway of the grid, and the tolerance distance for water molecules to be considered in the calculation (the ‘shell’ distance). The accuracy of our algorithm can also be increased by averaging density measurements over a number of rotations of the protein-water system in relation to the grid. We  evaluated the accuracy and computational cost of different combinations of parameters (leeway, grid size, shell distance, number of rotations, see SI and Figures S3, S4, and S5). We found that rotating the protein-water system was less computationally efficient than reducing the step size. Thus, we used a step size of 0.5 Å, shell distance of 6 Å, and a leeway of 5 Å, without any system rotation (see SI for details).

\subsection{Residue volume calculation}
\noindent
Similarly, it is possible to calculate the volume of an individual residue of a protein by determining which voxels are closer to the atoms of this residue than to the atoms of any other part of the system (i.e., the atoms of all other residues of the protein and the atoms of the solvent). The sum of these residue volumes for a single protein should be similar to the protein's volume, as calculated previously, with slight discrepancies due to the non-global nature of this type of calculation possible. To reduce such discrepancies, the voxel grid was defined in the same way for the global protein volume calculation and for the residue volume calculation, with the relevant sections of the surrounding voxel grid assigned to each residue in the latter case. In practice, a small discrepancy of 0.024\% is observed (see SI for details).

\subsection{Excess protein volume}
\noindent
Characterising the volume of a protein by the volume which it displaces in solvent, specifically defined at the surface of the protein by the midway point between protein atoms and the nearest solvent atoms, necessarily results in a region of what we define as protein volume being outside of the van der Waals radii of either protein or solvent atoms. This region, surrounding the van der Waals surface of the protein, can be estimated by calculating the volume of voxels which are both outside the van der Waals radii of their nearest protein or solvent atoms and at the solvent-interface surface of the protein. This excess volume can then be optionally removed to give a protein volume that matches the van der Waals protein volume at the protein's surface.

\subsection{Identification of buried waters}
\noindent
To identify waters that are buried inside the protein (which would typically not be considered part of the protein, but are likely to be an integral part of it) ~\cite{Park2005}, we used the DBSCAN clustering algorithm ~\cite{Ester1996} as implemented in the scikit-learn Python library ~\cite{scikit} to cluster water molecules by the coordinates of their oxygen atoms. Using a minimum sample size of 1 and 4.0 Å as the maximum distance allowed between samples within the same cluster, the first cluster (sorting by cluster size) represents the bulk water, including the hydration shells. Therefore, further clusters of water molecules tend to describe water molecules buried inside the protein, either alone or in groups. By filtering water molecules identified as being part of these clusters by ensuring that they are located within 3 Å of a protein atom reliably identifies water molecules that are buried within the protein. Optionally, our method enables calculating the protein density by considering internal water molecules as part of the protein.

\subsection{Analysis of water density}
\noindent

The density of the protein-solvent system can be calculated while successively including more solvent moving out radially from each protein atom, with all solvent atoms within a specific distance from the centres of each protein atom becoming part of a protein-solvent complex and thus considered in the density calculations (see Figure \ref{fig:method}). By excluding the protein's mass and volume contributions to this calculation, it is possible to calculate water density as a function of distance from the protein-water interface.

\subsection{Protein physical properties determination}
\noindent
We extracted various protein physical characteristics to evaluate any correlation between them and either the protein mass density, or the change in protein mass density upon including the effects of buried waters. These characteristic values include SASA, sphericity, aspect ratio ~\cite{Shannon2019}, amino acid composition, and protein molecular weight. 
The SASA is calculated via the Shrake-Rupley algorithm ~\cite{Shrake1973}, otherwise known as the “rolling ball” method. A probe size of 1.4 Å is usually used to represent a water molecule, an approximation of the water molecule’s van der Waals radius (more accurately, half of the oxygen-oxygen distance between two hydrogen-bonded water molecules ~\cite{Chang2006}). The amino acid compositions of surface and interior (i.e., non-surface) parts of the proteins were also calculated, with protein atoms being defined as surface atoms if more than 5\% of the spherical mesh points surrounding each atom were found via the SASA algorithm to be surface accessible to a 1.4 Å probe. \par
We assessed the normality of distribution of protein density and each physical characteristic for normality using the Shapiro-Wilk Test for normality ~\cite{Shapiro1965}. We calculated Pearson’s correlation coefficients between the protein mass density (which was found to be approximately normally distributed) and physical characteristics. As not all physical characteristics were found to be normally distributed, we also evaluated the Spearman correlation coefficient, which is more appropriate in these circumstances (see Tables S1 and S2).
\par

\subsection{Random Forest Regressor}
\noindent
To predict protein densities, we trained random forest regressors (RFR) using the scikit-learn Python library ~\cite{scikit}. We built two random forest regressors: one trained using only sequence-derived features (seq-RFR) and another trained using sequence and structural features (struct-RFR). The features included in the latter included amino acid residue composition, secondary structure (helix, strand, and coil percentages), net charge, and the prevalence of hydrophobic residues.
We optimised the regressors' hyperparameters (maximum depth, minimum number of samples per leaf, and number of estimators) via a grid search with cross-validation, measuring the performances of cross-validated models by their $R^{2}$ score to ensure the residuals are minimized. For struct-RFR, we also carried out an ablation study to identify the minimal set of features leading to highest performance in the classifier, which yielded a classifier operating on 20 features. For details on selected features, and training and validation protocols, see Supplementary Methods, Tables S1 and S2, and Figures S17-22.

\subsection{Markov State Modelling of BPTI}
\noindent
A down-sampled version of the 1-millisecond BPTI simulation performed by DE Shaw et al. ~\cite{Shaw2010}, with a timestep of 10 ns, was used to determine distinct conformational states of the protein via a Markov State Modelling (MSM) analysis using the PyEMMA Python package ~\cite{pyemma}. \par
Dimensionality reduction was performed using Time-lagged Independent Component Analysis (TICA) ~\cite{tica1, tica2} of the $\alpha$-carbon Cartesian coordinates. The TICA coordinates were subsequently clustered using $\kappa$-means clustering ~\cite{kmeans} to produce 300 discrete clusters. We verified the estimated Markov State Model via an well-established procedure ~\cite{prinz2011probing}. First, by analysing its implied timescales, and by performing a Chapman-Kolmogorov test (see Figure S23). Perron-Cluster Cluster Analysis ++ (PCCA++) ~\cite{pcca, pcca2} algorithm is then used to assign a probability for each $\kappa$-means state being a member of a smaller collection of 5 metastable macrostates. We then find the $\kappa$-means cluster with the highest probability of being in each metastable PCCA++ state and sample 50 trajectory frames that are associated with each, for a total of 250 protein conformations over 5 metastable states. All atom protein conformations were then solvated with TIP3P water, simulated with the Amber ff14SB force field, and had their density calculated via the routine previously detailed for the 260-protein dataset.

\subsection{Steered molecular dynamics of titin}
 \noindent
The I27 domain of titin (PDB: 1TIT) was aligned so that the vector connecting N- and C-terminus lays on the x-axis, and solvated in TIP3P water (155x60x62 \AA). The resulting box was neutralised with Na$^+$ counterions, and simulated with the Amber ff14SB force field, using a 2 fs time step, and PME handling long range electrostatic interactions. The system was first energy minimised for 500 steps, then simulated for 50 ps in the NVT ensemble, with 300 K imposed via Langevin dynamics (damping of 5 ps$^{-1}$) and $\alpha$-carbons restrained with a harmonic potential of 10 kcal mol$^{-1}$. Maintaining the restraints, 100 ps were then simulated in the NPT ensemble with 1 Atm imposed by a Langevin piston (period of 200 fs, decay of 50 fs), followed by 1 ns without restraints. From this equilibrated state, the unfolding of titin subject to mechanical stress was simulated via steered molecular dynamics (SMD). To this end, the N-terminus of titin was restrained with a harmonic potential, while the C-terminus was pulled for 8 ns at a constant velocity of 25 \AA~ns$^{-1}$ along the vector connecting the termini (thus extending within the elongated water box), with a force constant of 7 kcal mol$^{-1}$ \AA$^{-2}$.

From the SMD, we extracted titin conformations every 20 ps between 1 and 7 ns, for a total of 291 titin conformations. To ensure water is correctly packed around each extracted conformation, removing effects that might be caused by a fast steering or imposed restraints, each system was re-solvated and relaxed for 1 ns without any restraint, utilising the procedure previously outlined for the 260-protein dataset. For each of these individual simulations, We extracted snapshots at 0, 0.5 ns and 1 ns, for a total of 873 solvated titin conformations. For each of those conformations, we calculated density and percentage of amino acids part of a $\beta$ strand calculated in MDTraj ~\cite{mdtraj} using an implementation based on DSSP-2.2.0 ~\cite{dssp}. We used the statistical distribution of this latter quantity, bimodal in nature, to subdivide the ensemble of titin conformers in two sub-populations using the minimum between its two peaks as classification criterion.

\section{Results}

\subsection{Mean protein density}

\noindent For each of the 4221 protein-solvent conformation extracted from each of 260 simulations of proteins in water, we identified internal water molecules for 88.2\% of structures (85.5\% when averaging over multiple simulation frames), with an average of 14 ± 8 buried water molecules identified. However, including these buried waters in our calculations had a statistically insignificant effect on the calculated protein density (see Figure S8). This is explained by the small volume occupied by a single water molecule (30 Å\textsuperscript{3}) ~\cite{McIntosh1986}, and by the number of internal waters varying linearly in proportion to the molecular weight
At 300 K, we calculated a mean protein density of 1.294 $\pm$ 0.004 g cm\textsuperscript{-3} when including the effects of internal water molecules, and 1.296 $\pm$ 0.001 g cm\textsuperscript{-3} when neglecting them. Our benchmarks also showed that changing the simulation temperature with the Amber14SB force field and TIP3P water model combination had only a minimal effect on the calculated densities (mean densities of 1.295 $\pm$ 0.002 g cm\textsuperscript{-3} at 300 K and 1.293 $\pm$ 0.002 g cm\textsuperscript{-3} at 310.15 K). Conversely, altering the simulation temperature yielded significant effects when using the Amber 99SB-ILDN force field and SPC/E water model combination, whereby at 310.15 K the mean protein density decreased to 1.289 $\pm$ 0.002 g cm\textsuperscript{-3} from 1.298 $\pm$ 0.0004 g cm\textsuperscript{-3} at 300 K (see Figures S9 and S10).

\subsection{Correlation of density with other physical properties}

\noindent While we did not find any correlation between protein mass density and protein mass (PCC: 0.02446, p-value: 0.7265, SCC: $-$0.04361, p-value 0.5326), many physical characteristic values of proteins did have small but significant correlations with the protein mass density (see Figures S11-S15). The overall protein charge, the percentage of charged residues, and the percentage of hydrophobic residues were found to have weak correlations with the protein mass density, with slightly larger correlations resulting from considering the amino acid composition of the surface and interior of the protein separately (SCCs ranging from $-$0.1672 to $-$0.5142). \par
Aggregating the mass densities of all the protein-solvent conformations in our dataset we found a mean value of 1.289 $\pm$ 0.002 g cm\textsuperscript{-3}. The inclusion of buried water molecules in the density calculation reduced the calculated density for all protein-solvent conformations for which buried waters were identified, though the overall effect was so small that the sample mean remained unchanged. For proteins with multiple conformers, the mean standard deviation of the protein mass density was 0.006 g cm\textsuperscript{-3}, suggesting that dynamics in the sub-ns timescale had a limited effect on the protein mass density. While fast dynamics have limited effect on protein volume (and hence protein mass density), slower dynamics over longer timescales might still have a significant effect.

\begin{figure}[h!]
  \includegraphics[scale=1.0]{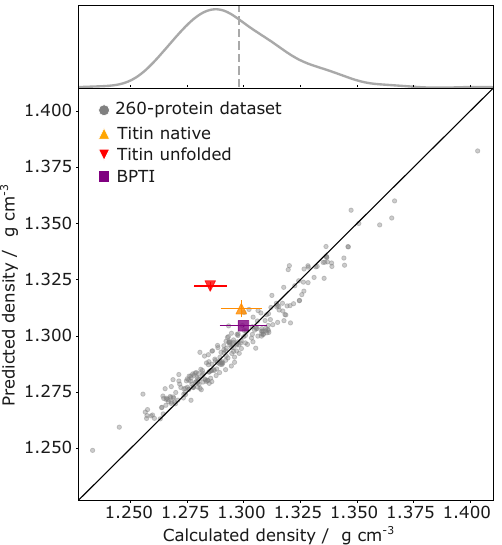}
  \caption{Comparison between calculated densities, and densities predicted by a Random Forest Regressor (struct-RFR). The RFR was trained on a set of 20 sequence- and structure-based features gathered from our 260-protein dataset, and the equilibrated crystal structure of BPTI (PDB: 5PTI) and titin (PDB: 1TIT) in its folded and extended state.  The identity line is shown in black for comparison. We find a PCC of 0.976 for the 260-protein dataset. The RFR accurately predicts the densities of conformations of BPTI (of which one structure is present in the training set) and folded titin (not part of the training set). The density of mechanically unfolded titin conformations are poorly predicted. Above, a Kernel Density Estimation plot of the distribution of mean calculated densities in the 260-protein dataset, with the overall mean density of 1.296 $\pm$ 0.001 g cm\textsuperscript{-3} annotated with a dashed vertical line.}
  \label{fig:rf}
\end{figure}

\subsection{Protein density prediction}
\noindent 
We investigated whether the density of 
 a protein can be predicted based on a collection of structural and sequence features. To this end, for each protein we measured 40 structural features, and amino acid composition (20 features quantifying the percentage presence of each amino acid in the protein sequence). We then trained Random Forests regressors (RFR) with a combination of the structural and sequence features (struct-RFR), or with sequence features alone (seq-RFR). In both circumstances, we found that the regressors were able to accurately predict protein densities, with seq-RFR (mean squared error (MSE): 3.88e-0.5, Pearson correlation coefficient (PCC): 0.967) outperforming struct-RFR (MSE: 2.19e-05, PCC = 0.981). This means that protein densities can be quickly and accurately predicted from only a protein’s amino acid sequence. Reviewing the importance of each feature, we found that protein mass is one of the least important features (see Figure S21). The most important features for struct-RFR are the percentage of aliphatic hydrophobic residues, the percentage of hydrophobic residues, and the total protein charge.

\subsection{Density variation in individual proteins}
\noindent
To evaluate the extent of density variations within individual proteins, we studied two  different cases, a long equilibrium simulation of BPTI, and a non-equilibrium simulation reproducing the unfolding of an Ig-like domain of titin under mechanical stress. BPTI is featured in the training set, whereas the titin domain is not (see Figure S2).

\begin{figure}[h!]
  \includegraphics[scale=1.0]{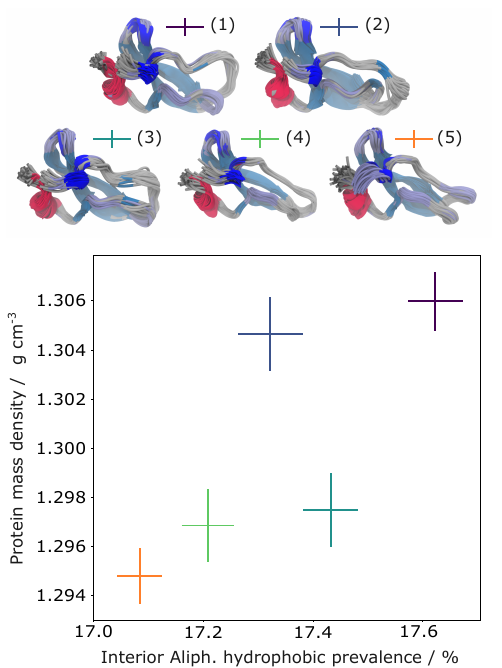}
  \caption{BPTI features states with distinct densities. A 1 ms molecular dynamics simulation of BPTI can be subdivided in five metastable states via Markov State Modelling (each represented by 50 overlaid conformers at the top). The secondary structure is coloured as: $\alpha$-helix in red, $3_{10}$-helix in blue, $\beta $-sheet in turquoise, turns in light blue and coil in grey. These conformers differ in their degree of compaction, as captured by the prevalence of aliphatic hydrophobic residues in their interior and their density. The structures of states 4 and 5 most closely resemble the crystal structure of BPTI (PDB: 5PTI), with a difference of only a more significant turn in the coil region of residues 12 and 13 in state 5. States 1 and 2 are differentiated from the others by the loss of structure of the small N-terminal ${3_{10}}$-helix, though for some conformers this feature is intact.}
  \label{fig:bpti}
\end{figure}

For BPTI, we found that the simulation is divisible into five metastable states with distinct densities and interior aliphatic hydrophobic residue prevalences (see Figure \ref{fig:bpti}). These are associated with structural differences in the N-terminal $3_{10}$ helix --- which is less prevalent in the conformers of states 1 and 2 --- and different arrangements of the loops between residues 7-16 and 35-46, leading to different degrees of compaction. 

\begin{figure}[h!]
   \includegraphics[scale=1.0]{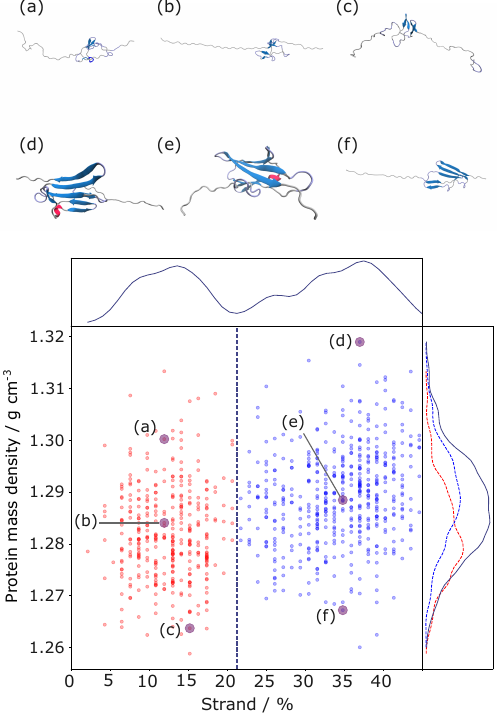}
  \caption{Unfolding of titin under mechanical stress reveals conformer sub-populations with distinct densities. For each titin conformation in a steered molecular dynamics simulation (see example conformers at the top), we calculate density and percentage of preserved secondary structure. The distribution of secondary structure content (kernel density estimation in the upper graph) reveals two distinct sub-populations, identified with red and blue colours in the scatter plot. In the right graph, kernel density estimations reveal that these sub-populations feature distinct density distributions (red and blue lines). The density distribution of the whole simulation is shown in black.}
  \label{fig:titin}
\end{figure}

\newpage
For titin, we found that the 873 equilibrated conformations extracted from the SMD simulation could be separated into two distinct sub-populations based on the percentage of residues found in a $\beta$-strand conformation. The sub-population associated with lower strand percentage (i.e., the more unfolded conformations, red in Figure \ref{fig:titin}) has a mean density of 1.2819 $\pm$ 0.0005 g cm\textsuperscript{-3}. The sub-population associated with higher strand percentage (i.e., conformations closer to the native state, blue in Figure \ref{fig:titin}), has instead a marginally higher, though distinct (t-test statistic: -11.7, p-value: 1.61e-29), mean density of 1.2892 $\pm$ 0.0004 g cm\textsuperscript{-3}.

Importantly, while both BPTI and titin feature conformations of varying density, we found that these variations fall within the distribution of densities of proteins with comparable physical properties in our protein training set (see Figures \ref{fig:rf} and S2). We challenged our trained struct-RFR with snapshots from the simulations of BPTI and titin, subdivided in its folded (native) and unfolded (extended) subpopulations. Density predictions for BPTI conformers were accurate overall, titin native conformers were slightly overestimated, whereas extended titin conformations were incorrect (see Figure 2). This failure is not unexpected, given that the training set consisted only of proteins in their folded native state. Qualitatively, this failure is explained by the fact that the most important feature for struct-RFR is the prevalence of aliphatic hydrophobic residues interior, a quantity which is negatively correlated with protein density. Hence, when titin unfolds and the interior aliphatic hydrophobic residues become exposed to solvent, the internal prevalence of these residues decreases, leading the RFR to predict a lower density than we calculate.

\subsection{Effect of hydration on protein density}

\begin{figure}[h!]
   \includegraphics[scale=0.95]{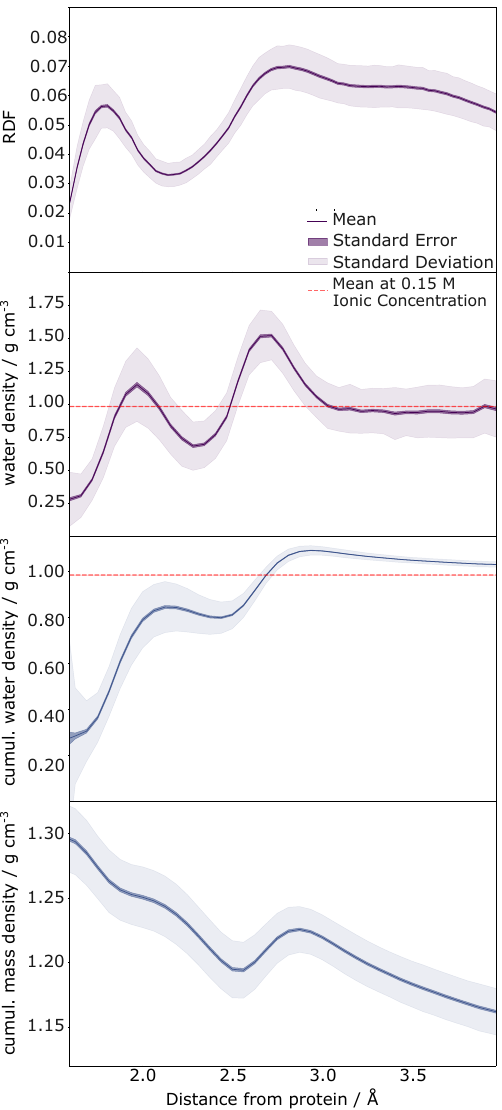}
  \caption{Relationship between protein and water density, averaged over the whole protein dataset. The two top graphs, in palatinate colour, report on water radial distribution function (RDF) and water mass density. The bottom two graphs, in blue, report on the cumulated effect on measured mass density of water, or the combined protein-water system, when accounting for an increasingly large water shell around the protein. The effective protein-water mass density decreases the more water is included, with a non-monotonical trend determined by water having density higher than bulk value in the first two hydration shells.}
\label{fig:rdf_density}
\end{figure}

\noindent Water forms hydration shells around solutes, with water molecules in contact with the protein featuring dynamics more akin to those of the protein, than those of bulk water ~\cite{laage2017water}. We therefore investigated how the measured protein density might be altered in situations whereby water molecules in the immediate vicinity of the protein are included in the calculation (see Figure \ref{fig:rdf_density} and S24). Averaging over all proteins in our dataset, we obtained a mean thickness of the first hydration shell of 2.197 $\pm$ 0.001 Å, with a standard deviation of 0.01 Å. Including the first hydration shell in our calculations led to a mean protein mass density of 1.228 g cm\textsuperscript{-3} , while including the second hydration shell (found at a cutoff distance of 3.322 $\pm$ 0.004 Å) led to a density of 1.185 g cm\textsuperscript{-3}. 
Interestingly, we observe that the density of a protein-water system decreases non-monotonically when an increasing amount of water surrounding the protein is included. This indicates that the density of water surrounding the protein is not constant.

\subsection{Hydration shell structure}
\noindent

Our results show that the presence of a protein molecule can significantly alter the density of the water that surrounds it. So, we finally quantified how the presence of a protein might affect adjacent water structure (see Figure \ref{fig:rdf_density}). We found that, for the first and second solvation shells, water is denser than bulk. Specifically, on average water reaches a density of  1.1 $\pm$ 0.3 g cm$^{-3}$ in the first shell (12\% greater than bulk water), and an even larger density of 1.5 $\pm$ 0.2 g cm$^{-3}$ in the second shell (54.5\% denser than bulk). Investigating the order of water at a range of distances from proteins (see Figure S26) we also observed that this becomes more organized with successive solvation shells. Considering the density of an increasingly large shell of water around a protein, we observe that if at least two water shells are present, the average water density will be greater than bulk, slowly converging to bulk if more water is accounted for.

\section{Discussion and Conclusion}
\noindent
In this work we have produced a large dataset of hydrated protein structures via molecular dynamics simulations to accurately assess the mass density of proteins, establish any possible correlations between mass density and physical characteristic values of proteins, determine the effect of the inclusion of hydration shells on the apparent protein mass density, and investigated the effect of the protein on the organisation of the water around it. To determine protein mass density, we developed and profiled an efficient voxel-based method, which is also able to identify and account for buried waters. \par
For our dataset, we calculated a protein mass density of 1.296 $\pm$ 0.001 g cm\textsuperscript{-3} at 300 K.
These measures are essentially unaffected by the presence of buried water molecules, and only marginally increased when using a different force field and water model. Overall, the values we measured are lower than the 1.35 g cm\textsuperscript{-3} value commonly used in the scientific literature. Furthermore we found that, in contrast to previous research, there is no correlation between protein density and mass. However, we identified other physical characteristics that are significantly correlated with the protein mass density. These include the overall charge, the percentage of hydrophobic amino acid residues, and the percentage of charged surface amino acids. We also demonstrated that these correlations can be exploited by a Random Forests regressor to predict with high accuracy the densities produced by our MD-based method, at a fraction of the computational cost. Remarkably, we also found that a regressor could yield high quality predictions based on the amino acid sequence alone.\par

As proteins are dynamic in nature, we also investigated how the density of an individual protein might evolve using molecular dynamics simulations of BPTI and titin. While our regression model demonstrated that the main determinant of density in a protein is amino acid composition, our results show that conformational changes also have a measurable effect. Given that such dynamics-dependent effects are subtle, we expect an amino acid composition-dependent density value to be a suitable proxy for most experiments. However, the variations we observed highlight the presence of volume changes at biologically relevant frequencies, which might be identifiable with techniques sensitive to fluctuations in the distribution of charges and electron density in an analyte. \par

Finally, we characterized how the hydration shells affect the resulting measured density of both protein and water. The non-monotonic decrease in the density of the protein–water system when increasing the number of water molecules included is due to the varying density of water in the first two hydration shells. Indeed, we found that the hydration shells have a density higher than the bulk, with the second shell exceeding the density of the first. We found that including the first hydration shell in the protein-water nanobioparticle resulted in a mean protein mass density of 1.252 g cm\textsuperscript{-3}, while including the second hydration shell further reduced the measured mass density to 1.192 g cm\textsuperscript{-3}. These values are relevant for any experiment studying the properties of proteins in solution. Overall, this observation highlights how considering water as bulk around an analyte of interest might affect the quantities extracted from a measurement. For instance in tasks such as background subtraction, e.g., in IR spectroscopy, whereby bulk water signal is removed from a spectrum to highlight the signal of an analyte in solution. In the context of protein density measurement, experimental techniques that estimate protein density often assume that water is a medium of constant density ~\cite{Gekko1979}. The density of a weakly hydrated protein might be overestimated if water is treated purely as bulk though, as part of the "extra mass" observed is explained by water being on average denser around the protein.

\begin{acknowledgments}

We wish to thank Durham HPC Hamilton for computational resources. We acknowledge the EPSRC and the SOFI2 CDT (grant EP/S023631/1) for financial support.
\end{acknowledgments}\noindent

\section{Author Declarations}

\subsection{Conflict of Interest}
\noindent
The authors have no conflict of interest to disclose.

\subsection{Author Contributions}
\noindent
\textbf{Cameron C. W. McAllister}: Conceptualization, Methodology, Investigation, Software, Formal Analysis, Visualization, Writing - Original Draft Preparation. \textbf{Lucas S. P. Rudden} Conceptualization, Resources. \textbf{Elizabeth H. C. Bromley}
Conceptualization, Methodology, Supervision, Writing - Original Draft Preparation. \textbf{Matteo T. Degiacomi}, Conceptualization, Methodology, Software, Investigation, Supervision, Writing - Original Draft Preparation.

\section{Data availability}
\noindent
Our method to calculate protein densities is available at \href{www.github.com/Degiacomi-Lab/ProteinDensity}{www.github.com/Degiacomi-Lab/ProteinDensity}.\\
The trained random forest regressors are available at \href{www.github.com/Degiacomi-Lab/DensiTree}{www.github.com/Degiacomi-Lab/DensiTree}.
The data that support the findings of this study are available from the corresponding author upon reasonable request.

\bibliography{density}
\end{document}